\documentclass{elsart}

\usepackage{natbib}
\usepackage{graphicx}
\usepackage{amssymb}

\newcommand{\EXPU}[3]{\mbox{\rm $#1 \times 10^{#2} \rm\:#3$}}  % exponent with units
\newcommand{\POW}[2]{\mbox{$\rm10^{#1}\rm\:#2$}}

\newcommand{\fuse}{\mbox{{\it FUSE}}}
\newcommand{\iue}{\mbox{{\it IUE}}}
\newcommand{\hst}{\mbox{{\it HST}}}

\def\VEL{\:{\rm km\:s^{-1}}}
\newcommand{\MSOL}{\mbox{$\:M_{\odot}$}}

\def\LA{Lyman\thinspace$\alpha$}

\begin{document}

\begin{frontmatter}



\title{Far Ultraviolet Spectroscopy of (Non-magnetic) Cataclysmic Variables}


\author{Knox S. Long}

\address{Space Telescope Science Institute,\\ 3700 San Martin Drive, Baltimore MD, 21218, U. S. A.}

\begin{abstract}
% Text of abstract
\hst\ and \fuse\ have provided high signal-to-noise,
high-resolution spectra of a variety of cataclysmic variables and
have allowed a detailed characterization of FUV emission sources
in both high and low states. Here, I describe how this has
advanced our understanding of non-magnetic CVs and the substantial
interpretive challenges still posed by the observations. In the
high state, the FUV spectra are dominated by disk emission that is
modified by scattering in high and low velocity material located
above the disk photosphere.   Progress is being made toward
reproducing the high-state spectra using kinematic prescriptions
of the velocity field and new ionization and radiative transfer
codes. In conjunction with hydrodynamical simulations of the
outflows, accurate estimates of the mass loss rates and
determination of the launching mechanism are likely forthcoming.
In quiescence, the FUV spectra reveal contributions from the WD
and the disk. Quantitative analyses have lead to solid
measurements of the temperatures and abundances of a number of WDs
in CVs, and of a determination of the response of the WD to an
outburst.
%Enough temperatures have been measured that
%differences between non-magnetic and magnetic CVs are being
%explored, leading to the suggestion that long-term averages of
%accretion rates are lower in magnetic CVs.
Basic challenges exist in terms of understanding the other
components of the emission in quiescence, however, and these are
needed to better understand the structure of the disk and the
physical mechanisms resulting in ongoing accretion in quiescence.
\end{abstract}

\begin{keyword}
accretion disks \sep  cataclysmic variables \sep dwarf novae \sep
white dwarf


\end{keyword}

\end{frontmatter}

% main text
\section{Introduction}
\label{sec_intro}

In disk-dominated cataclysmic variables (CVs), mass accretion onto
a white dwarf (WD) from a relatively normal secondary star is
mediated by a disk that extends close to the surface of the WD.
All CVs vary, but the character of the variability probably
reflects the time-averaged accretion rate from the secondary
\citep[see, e. g.][]{sthele01}. Systems with low accretion rate
show semi-periodic outburst of 3-5 magnitudes (in $m_v$) and are
known as dwarf novae (DNe).  These outbursts are due to a thermal
instability that converts the disk from a low temperature, mostly
unionized, optically thin (in the continuum) state to a high
temperature, ionized, optically thick state. During the outburst,
the mass transfer rate $\dot{m}_{disk}$ in the inner disk rises
from $10^{-10}$ to \POW{-8}{\MSOL~yr^{-1}}, and, in terms of the
simplest picture of the system, the dominant source of FUV
emission changes from being the WD to the disk. By contrast,
systems with high mass-transfer rates remain in the high state
most of the time, and are known as nova-like variables. Finally
there are ``Z Cam'' objects, which undergo normal DN outbursts as
well as outbursts that stall during the transition to quiescence
for weeks to months at an intermediate magnitude (typically 0.5-1
magnitudes below peak). These are thought to be systems with
intermediate rates of mass transfer from the secondary star.

In this review, I would like to highlight recent progress toward
understanding these systems as a result of observations with
\fuse, \hst, and the Hopkins Ultraviolet Telescope (HUT).

\section{Modelling the Disks and Winds of High State CVs}
\label{sec_histate}

\begin{figure}
\begin{center}
\includegraphics[angle=90,width=-2.5in]{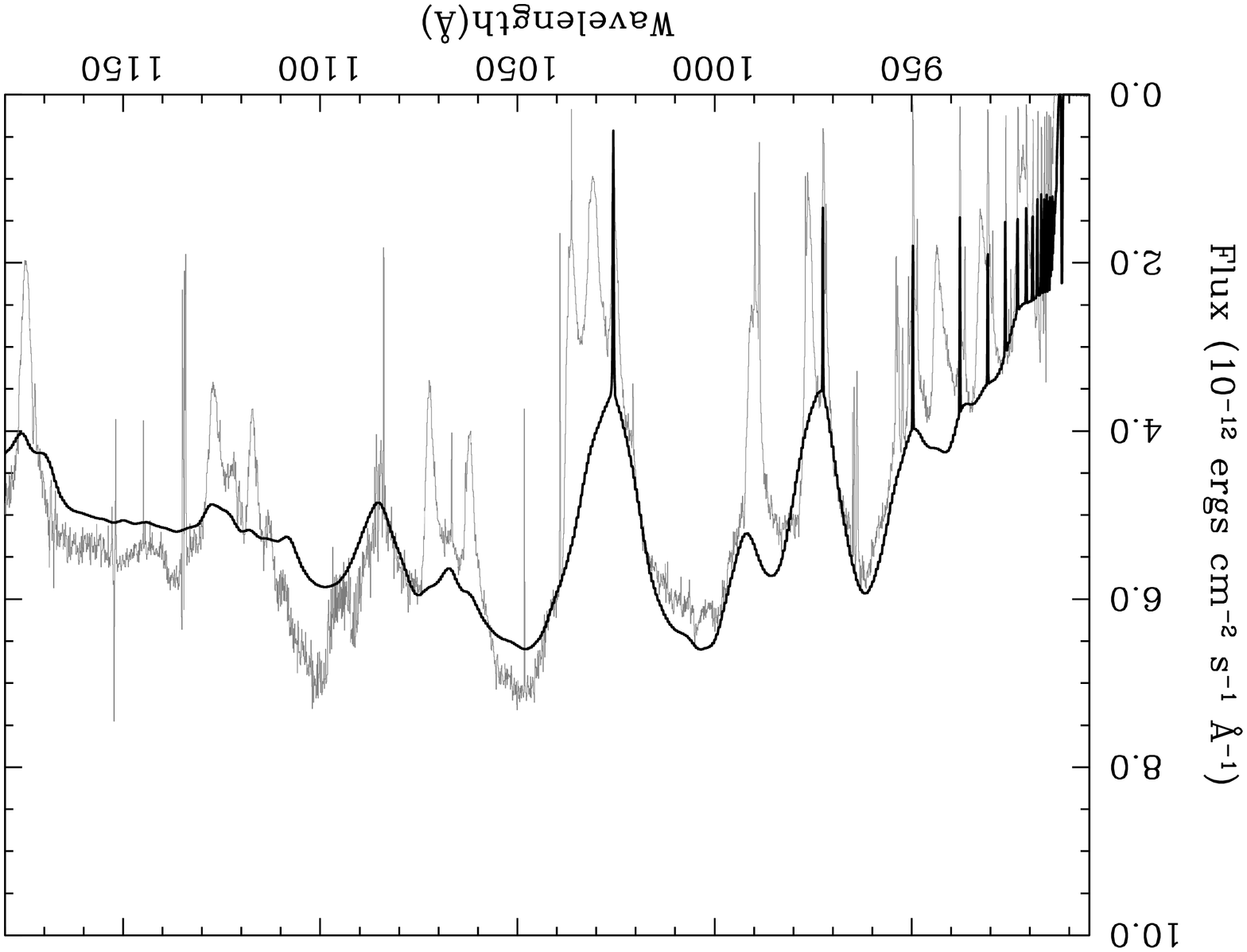}\includegraphics[angle=-90,width=2.5in]{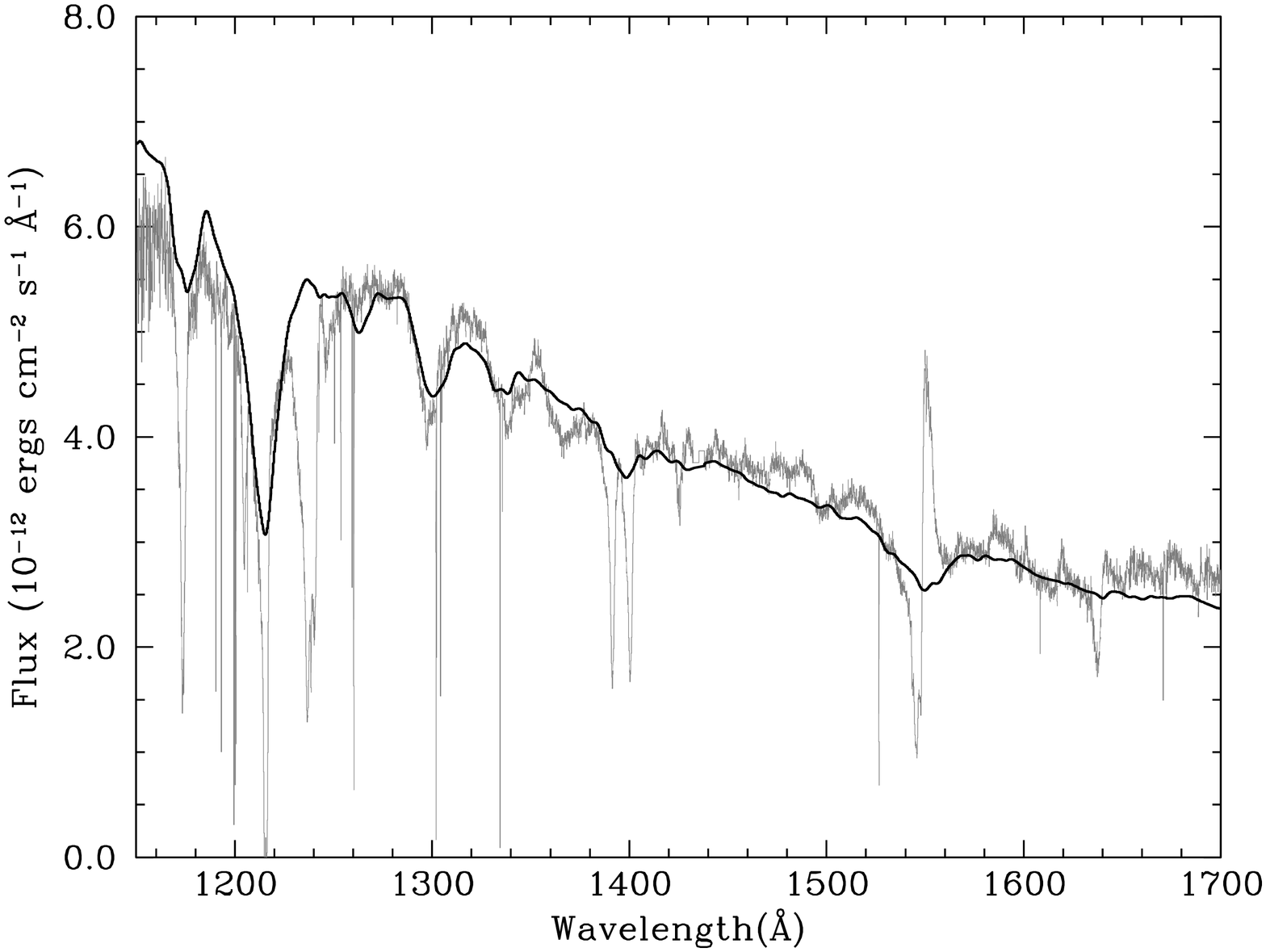}
\end{center}
\caption{V3885 Sgr as observed with \fuse\ (left) and \hst\
(right) compared to disk models constructed from summed stellar
spectra.  In fitting the models, the only free parameter was
$\dot{m}_{disk}$.  The distance to the object and the inclination
was fixed.  \label{v3885sgr_disk}}
\end{figure}

As first shown using \iue, the FUV spectra of disk-dominated CVs
in outburst show ``blue'' continua that can be reasonably
approximated in terms of a weighted set of stellar atmospheres,
where the weighting is determined by the temperature and gravity
of a steady state accretion disk and where each spectrum is
broadened to mimic that of a rotating disk annulus.  A modern
example using data from \fuse\ and \hst\, is shown in Fig.\
\ref{v3885sgr_disk}. When, as in this case, the distance and
inclination are known, the fit depends almost entirely on the
observed flux. In these cases, the fact that the model
qualitatively reproduces many of the features in the spectrum
suggests that the derived $\dot{m}_{disk}$ is close to the correct
value.  On the other hand, models based on stellar atmospheres
have generally been shown to fail to reproduce the spectrum over
large wavelength ranges, particular if the range spans the Balmer
limit. Stellar atmospheres have a pronounced Balmer jump in
contrast to what is observed in high state disk systems. This is
surely due to differences between disk and stellar atmospheres. Ad
hoc solutions, such as including a transition region above the
surface of the disk, are capable of addressing the problem but do
not discriminate between likely mechanisms, which include viscous
energy dissipation in or illumination of the disk atmosphere. To
date there have been very few published attempts to create more
physically correct model disk spectra and perform detailed
comparisons to high quality spectra. However, this is likely to
happen soon. The machinery does exist \citep{kriz86} and model
grids are beginning to be created \citep[see, e. g.][]{98wade}.

The disk model fits will not provide a complete solution, however,
because FUV spectra of high state spectra show clear evidence of
winds.  P Cygni like profiles in N V, Si IV and (most commonly) C
IV are observed in some systems, and the centroids of these lines
are blue-shifted in others. Blue edge velocities of 2000-5000
$\VEL$ are observed. Recently, a number of systems have been
observed in the 900-1185 \AA\ range with \fuse. These show
surprising diversity. In particular, although some systems do show
high-velocity wind emission from S VI and O VI, it is common to
see relatively narrow lines with small (100 $\VEL$) blue-shifts
from intermediate ionization states
\citep{froning01_ugem,froning2004_fuse_survey}. The velocity
widths of these lines are too narrow for either the disk or the
high-velocity outflow represented by C IV in the \hst\ range.

Shortly after winds were first discovered in CVs, observations of
eclipsing systems showed changes in profiles shapes that indicate
substantial rotation, suggesting a disk origin for the outflow
\citep{cordova1982,drew1988}. Consequently, our basic picture of
the high-velocity winds is of a bi-conical flow emanating from the
inner disk. \citet{93vitello} were the first to attempt to model
the profile shapes of wind lines as observed in high state CVs in
terms of a kinematic prescription for a bi-conical wind. They
found that the \iue-derived (R=200) C~IV profiles of three systems
-- RW Sex, RW Tri, and V Sge -- could be reproduced with
moderately collimated winds with mass-loss rates
($\dot{m}_{wind}$) of order 10\% of the disk accretion rate
($\dot{m}_{disk}$) and terminal velocities of 1-3 times the escape
velocity at the footpoint of each streamline. Subsequently,
\citet{97knigge} succeeded, using a different parametrization for
a bi-conical flow, in reproducing the C~IV profile of UX UMa
though an eclipse as observed at R=2000 with \hst/GHRS. This
analysis was the first attempt to model changes in the profile
through eclipse, and suggested, at least in UX UMa, the existence
of a relatively dense, high-column-density, slowly outflowing
transition region between the disk photosphere and the fast moving
wind.

\begin{figure}
\begin{center}
\includegraphics[width=3in,angle=-90]{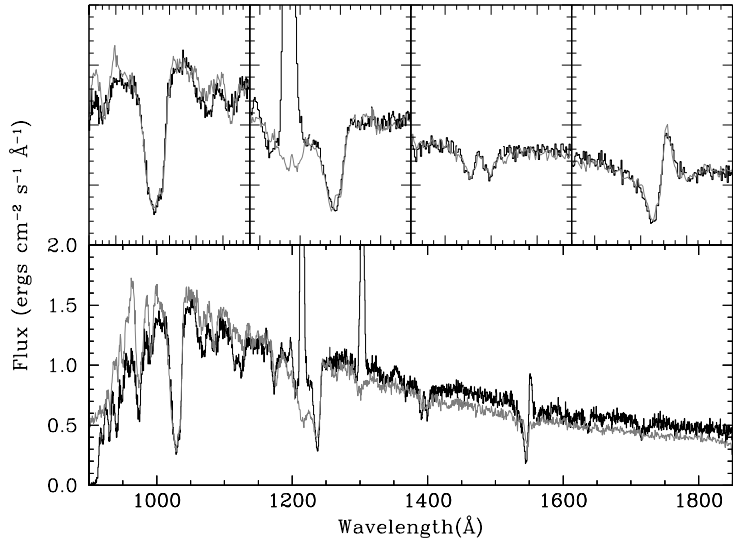}
\end{center}
\caption{Spectra of IX Vel as observed with HUT (black) compared
to synthetic spectra (grey) calculated with the Monte Carlo
radiative transfer code Python \citep{02long}. The upper panels
show from left to right best-fit comparisons to the O VI, N V, Si
IV and C IV regions. The lower panel wind shows the best model
when all of the line regions are fit simultaneously.
\label{ixvel_hut}}
\end{figure}

Recently, my collaborators and I have developed a new Monte Carlo
radiative transfer code that invokes a Sobolev approximation with
escape probabilities to follow photons through an axially
symmetric wind. ``Python'' is designed to produce a complete
spectrum of disk dominated CVs \citep{02long}. Emission sources
include the disk, the WD, a boundary layer and the wind itself.
The flow geometry is defined either in terms of the
parametrization of \cite{93shlosman} or \citet{95knigge}. The code
consists of two separate Monte Carlo radiative transfer
calculations.  In the first, the ionization structure of the wind
is calculated using a modified on-the-spot approximation. In the
second, a detailed spectrum is calculate for a specific wavelength
range of interest. A fair summary of progress to date is as
follows:  MC methods, such as Python, can be used to obtain
spectral verisimilitude to individual lines fairly easily. An
example of this is shown in Fig. \ref{ixvel_hut}. Systematic
searches are now needed to determine whether one can actually
model all the lines with a single wind geometry, and to determine
how well one can do on average. Several of us are beginning this
effort.

If this effort is successful, we will be able to determine the
basic parameters, such as the mass-loss rate, degree of
collimation, and ionization state, of the winds of high state CVs.
Until this is done, it is difficult to address many of the
physical questions that need to be answered, including whether the
wind is radiatively-driven, as is usually assumed. The
observational and theoretical evidence for this is murky. If the
wind is radiatively driven, one would expect that the wind lines
would be strongest when systems are brightest.  But
\citet{hartley02} found no correlation between the strength of
wind features and continuum brightness in  \hst/STIS observations
of two nova-like variables, IX Vel and V3885 Sgr. And
\citet{99feldmeier} and \citet{drew00} have argued that the
luminosities of CV disks are at best marginally sufficient to
accelerate a high-velocity wind. Hydrodynamical simulations of
radiatively-driven CV winds have recently been carried out, but
comparisons of the theoretical predictions to observations have
produced mixed results \citep[see e.g.][]{proga03}. Thus,
alternative wind-driving mechanisms must be (re)considered instead
of, or in addition to, radiation pressure. These include viscous
heating of the upper disk atmosphere \citep{czerny1989a}, and
irradiation of the disk \citep{czerny1989b} as well as
magneto-centrifugal forces leading to constant angular velocity
rotation out to the Alfv\`en surface \citep{cannizzo88}. Knowledge
of the kinematic structure of the wind would be a strong
discriminant between them.

\section{FUV emission of non-magnetic CVs in quiescence}
\label{sec_lostate}

\begin{figure}
\begin{center}
\includegraphics[width=2.5in]{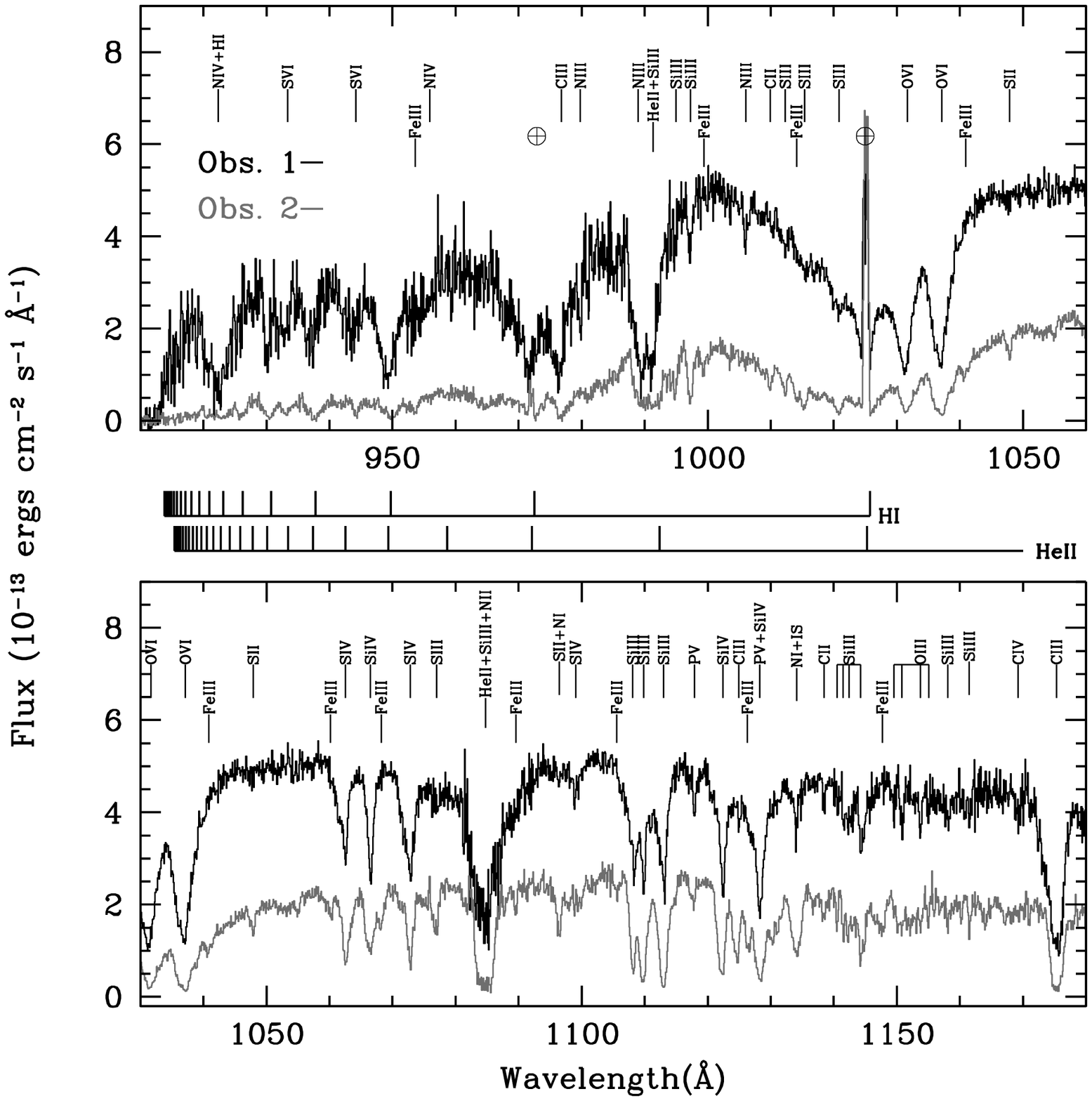}\includegraphics[width=2.5in, height=2.5in]{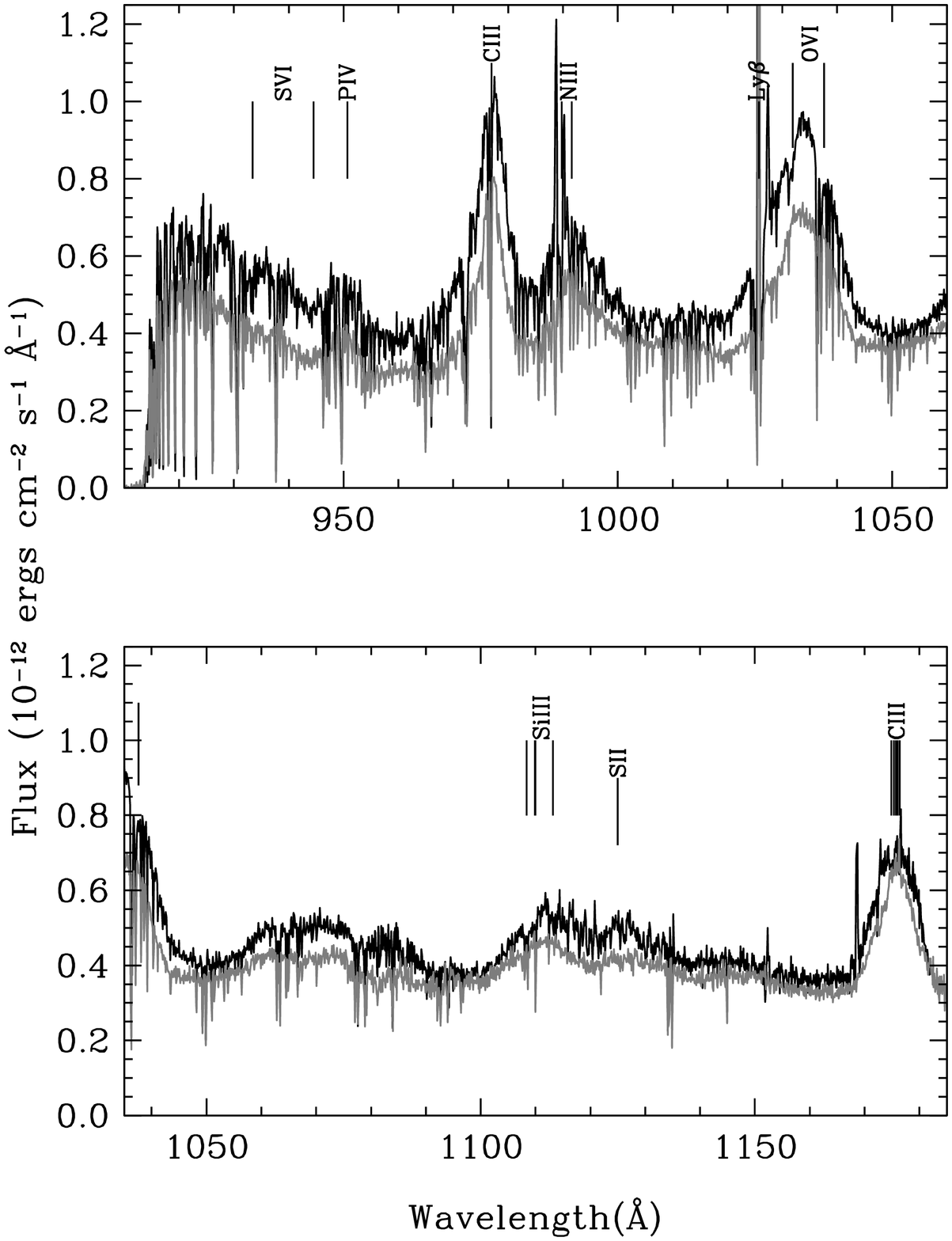}
\end{center}
\caption{U Gem and SS Cyg as observed with \fuse.
 Two spectra of each are plotted, one at the beginning of a quiescent interval and one in mid-quiescence. In U Gem,
 the difference is due to the cooling of the WD from $T_{wd}$ of about 40,000K to 30,000K. \label{ugem_sscyg}}
\end{figure}

The appearance of quiescent systems is quite different from those
in outburst, and fairly varied.  At one extreme are systems, such
as U Gem, with \fuse\ spectra (Fig. \ref{ugem_sscyg})
characterized by broad Lyman absorption lines and narrower metal
absorption features. These systems can be modelled, sometimes in
great detail, with synthetic spectra of metal enriched
WD-photospheres. The metal enrichment is understood to be a result
of on-going accretion. At the other extreme are systems dominated
by broad, fairly symmetric emission features. \fuse\ spectra of SS
Cyg, also shown in Fig.\ \ref{ugem_sscyg}, provide a good examples
of this type a system. The dominant continuum emission in these
systems is not well-understood, although the lines almost surely
arise from the disk \citep[see, e. g.][]{long05_sscyg_wxhyi}.
Since in quiescence the bulk of the disk is cold and unionized,
explanations typically involve surface layers of the disk or a
chromosphere, heated via X-ray illumination or magnetic flares in
the disk. The emission region is unlikely to be optically thick in
the continuum since that would limit the physical size of the
emitting region to that of the WD. Most systems, e.g. VW Hyi
\citep{godon04_vwhyi}, lie between the two extremes, with
contributions from the WD and from the ``second component''.

The basic questions one would like to answer about quiescent
systems depends on the nature of the emission.  For the WD
systems, one would first of all like to establish the fundamental
parameters -- temperature, radius, abundances, rotation rate,
gravitational redshift -- of the WD.  In addition, one would like
to explore the effects of accretion onto the WD.
%How much and for
%how long are the surface temperatures of the WD heated by the
%outburst?  Does the outburst create a long-lived rotating
%accretion belt on the surface?  How does accretion affect the
%interior temperatures in the long term?
For the other systems,
one would like to understand the real nature of the ``second
component'' and what it implies about the structure of the disk
(and the nature of the WD).

\begin{figure}
\begin{center}
{\includegraphics[origin=c,angle=-90,totalheight=2.5in]{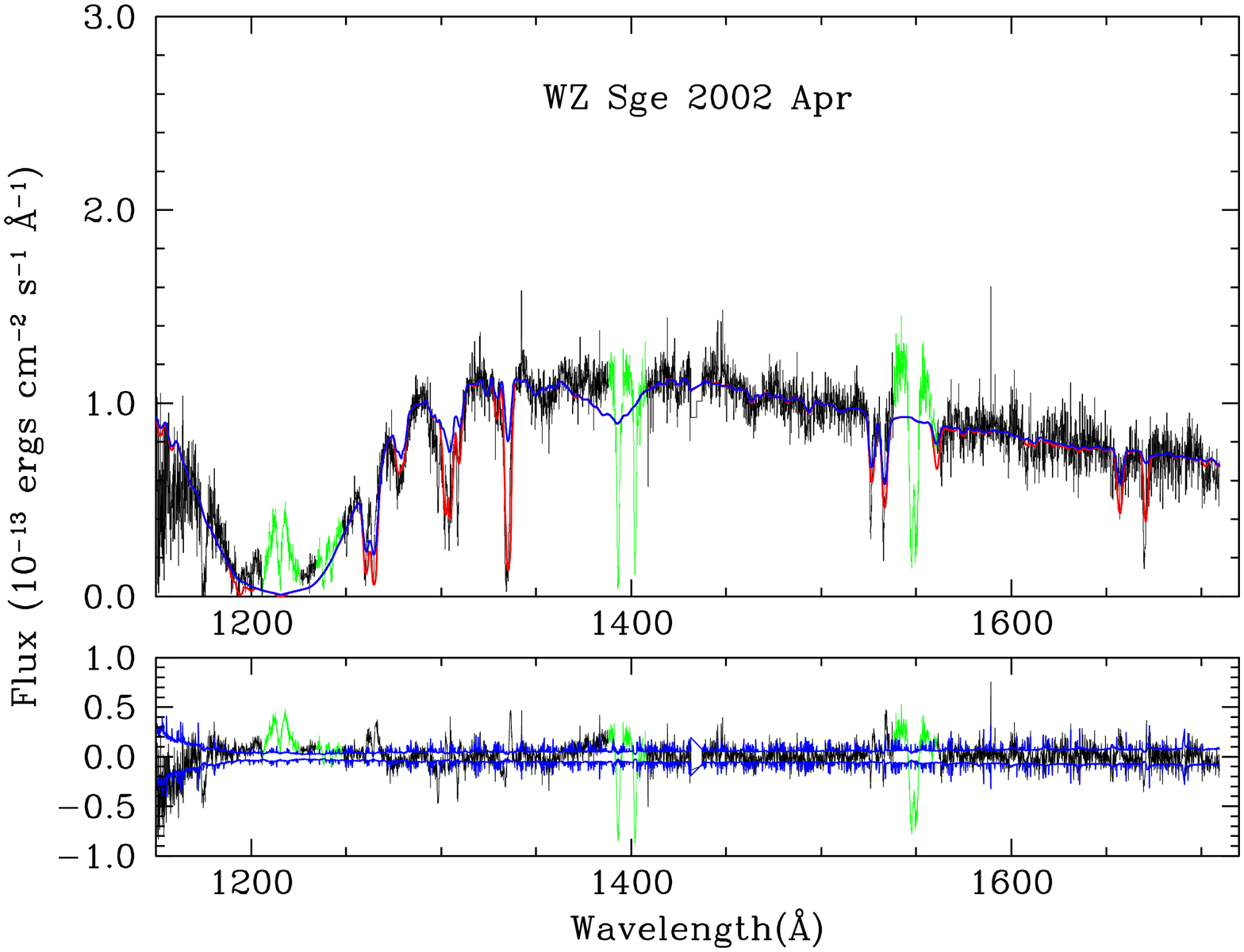}
\includegraphics[totalheight=2.5in]{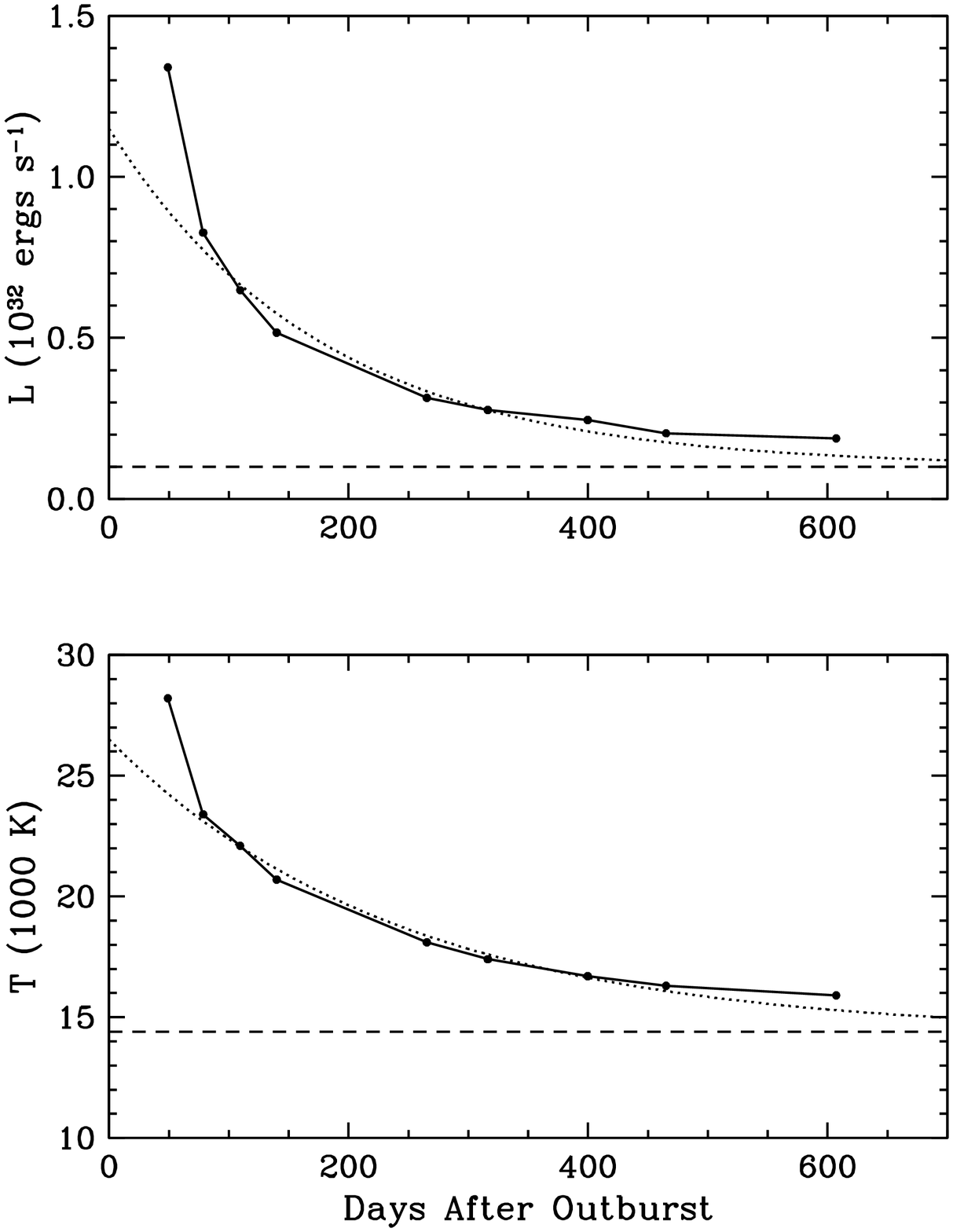}}
\end{center}
\caption{The left panel is the April 2002 \hst\ spectrum of WZ Sge
fit to synthetic spectra generated for a metal-enriched WD
photosphere.  The right panel shows the evolution of luminosity
and $T_{wd}$ following the 2001 outburst
\citep{long04_wzsge_stis}. \label{wzsge}}
\end{figure}

As an example, I would like to describe the analysis of multiple
FUV spectra that were obtained of WZ Sge following its outburst in
July 2001 \citep{sion2003_wzsge_stis,long04_wzsge_stis}.  WZ Sge,
at 42 pc, is the closest known CV, and possibly the closest
accretion disk system. Among CVs, WZ Sge has one of the shortest
binary periods (88 m), and one of the longest interoutburst
periods (20-30 yrs). \fuse\ and \hst\ observations during the
outburst showed a high excitation spectrum dominated by broad O
VI, NV, and C IV features, consistent with the high (75$^o$)
system inclination \citep{long03_wzsge_fuse,knigge02_wzsge}.
Following the outburst, the spectrum acquired the characteristics
of a WD-dominated system (Fig.\ \ref{wzsge}). From 2001 September
to 2003 March, the spectra show a steady decline in the flux, a
steady decrease in the color temperature,and a gradual increase in
the width of \LA, all of which are consistent with a long term
cooling of a WD.  A detailed analysis shows that $T_{wd}$ declined
from 24,300 in 2002 October to 15,900 K in March 2003. The flux
level in 2003 was still well above that of the pre-outburst
system, indicating that the WD was still cooling. Assuming a
normal WD mass-radius relationship and a distance of 42 pc, it was
possible to show that all of the post-outburst spectra are
consistent with a WD mass of 0.9\MSOL, as had been inferred
previously by more indirect means.  This result is interesting in
part, because in some other systems (e.g U Gem), simple 1-T models
do not produce constant radius with time, which has lead to
suggestions of non-uniform surface temperatures of the WD or a
declining ``second component''.

%The strength of the metal lines also appears to have declined,
%although the interpretation is a little less clear.  One
%possibility is that the lines are photospheric. If that is the
%case, then the mean metallicity of the atmosphere dropped from 2x
%solar in 2002 October to 0.4 solar in 2003 March.  If this is
%correct, then

One of the puzzles in the case of WZ Sge is why the post-outburst
energy release has been so large; the post-outburst excess energy
release is roughly 15\% of the outburst energy. There are two
basic possibilities.  First, there could be on-going accretion.
This is consistent with the fact that we continue to see metal
lines in WZ Sge. The time scale for metals to settle in a WD are
of order a few days.  Therefore, if the metal lines in WZ Sge
arise from the photosphere, then there must be ongoing accretion.
Secondly, the WZ Sge outburst both heated and left the WD with
extra mass on its surface, leaving the WD out of equilibrium.  The
interior of the star has to respond to these facts, through a
process known as compression heating. \citet{godon2004_wzsge}
argued that the cooling that is observed in WZ Sge could be
modelled either as a 1.2 \MSOL WD which accreted at
\POW{-8}{\MSOL~yr^{-1}} during the outburst with no accretion
afterward, or with a 1 \MSOL WD accreting at
\EXPU{2}{-9}{\MSOL~yr^{-1}} during the outburst and
\EXPU{2}{-11}{\MSOL~yr^{-1}} following the outburst.  The latter
possibility is most consistent with other known facts about WZ
Sge.

\section{Conclusions}
\label{sec_conlustions}

The spectra obtained with \fuse, \hst, and HUT represent
qualitative improvements over those that could be obtained with
\iue.  For the high state spectra, the surprise in the \fuse\
spectra is the plethora of lower ionization state narrower lines,
especially in higher inclination systems.  These lines most likely
arise from a transition region between the disk and the high
velocity wind, another indication that this zone cannot be
ignored.  With \iue\, it was clear that WDs emission dominated the
emission from some low state spectra.  With \hst\ and \fuse, it is
possible to model not only the temperature evolution of WDs, but
to determine abundances in the photosphere and to begin to
characterize the second component in the emission. The existing
spectra demonstrate the importance of multiple observations of
individual systems. Sadly, unless a servicing mission provides
\hst\ with a revived UV spectroscopic capability or the \fuse\
operations staff are able to restore \fuse\ to operational status,
the prospects for obtaining new spectra are modest.

Acknowledgement: This work was supported by NASA through grants
associated with analyses of \hst\ and \fuse\ spectra of CVs,
specifically GO-9357 and GO-9791 from the Space Telescope Science
Institute, and NAG5-13706 and NAG5-13717 directly from NASA.

% \appendix

% \section{}
% \label{}

% Bibliographic references with the natbib package:
% Parenthetical: \citep{Bai92} produces (Bailyn 1992).
% Textual: \citet{Bai95} produces Bailyn et al. (1995).
% An affix and part of a reference:
%   \citep[e.g.][Ch. 2]{Bar76}
%   produces (e.g. Barnes et al. 1976, Ch. 2).

\end{document}